%% file: acl_latex.tex
\documentclass[11pt]{article}

\usepackage[preprint]{acl}

\usepackage{times}
\usepackage{latexsym}
\usepackage[T1]{fontenc}

\usepackage[utf8]{inputenc}

\usepackage{microtype}

\usepackage{inconsolata}

\usepackage{graphicx}
\usepackage{hyperref}
\usepackage{url}
\usepackage{algorithm}
\usepackage{algpseudocode}
\usepackage{algorithmicx}
\usepackage{xspace}        
\usepackage{enumitem}
\usepackage{array}
\usepackage{arydshln}   
\usepackage{makecell}   
\usepackage{pifont}    
\usepackage{wrapfig} 
\usepackage{multirow}
\usepackage{comment}
\usepackage{amsthm}           
\usepackage{natbib}
\usepackage{booktabs}
\usepackage{xcolor}
\usepackage{caption}
\usepackage{amsmath}
\usepackage{amssymb}
\usepackage{multicol}
\usepackage{cuted}
\usepackage{cuted}
\usepackage{capt-of} 
\usepackage{subcaption}
\usepackage{graphicx}
\usepackage{tikz}
\usepackage{tabularx}
\usepackage{listings}
\usepackage{bbding}
\usepackage{makecell}
\usepackage{dblfloatfix} 
\usepackage{placeins}    
\newcolumntype{Y}{>{\raggedright\arraybackslash}X}
\AtBeginDocument{%
	\renewcommand{\eqref}[1]{Eq.~\textup{(\ref{#1})}}%
}

\definecolor{AblationColor}{HTML}{006400} 

\newcommand{\projectname}{\textsc{MGTeval}\xspace}

\title{\projectname: An Interactive Platform for Systemtic Evaluation of Machine-Generated Text Detectors} 

\author{Yuanfan Li\textsuperscript{$1,\dagger$}, 
Qi Zhou\textsuperscript{$1,\dagger$}, 
Chengzhengxu Li\textsuperscript{$1,\dagger$}, 
Zhaohan Zhang\textsuperscript{$2,\dagger$},
\textbf{Chenxu Zhao}\textsuperscript{$1,\dagger$},  \\
\textbf{Zepu Ruan}\textsuperscript{$1,\dagger$}, 
\textbf{Chao Shen}\textsuperscript{$1$}, 
\textbf{Xiaoming Liu}\textsuperscript{$1,\ast$} 
\\
        \textsuperscript{1}Faculty of Electronic and Information Engineering, Xi'an Jiaotong University\\ 
        \textsuperscript{2}Queen Mary University of London \\
        \textsuperscript{$\dagger$} Equal contribution, \textsuperscript{$\ast$} Corresponding author: \texttt{xm.liu@xjtu.edu.cn}\\
        Project Website: \url{http://uncoverai.cn} \\
        Project Codebase:  \url{https://github.com/Liyuuuu111/MGT-Eval} \\
        Demo Video: \url{https://www.youtube.com/watch?v=1CVoGQFW4KU}
        }

\begin{document}
\maketitle

\begin{abstract}
We present \projectname, an extensible platform for systematic evaluation of Machine-Generated Text (MGT) detectors. 
Despite rapid progress in MGT detection, existing evaluations are often fragmented across datasets, preprocessing, attacks, and metrics, making results hard to compare and reproduce. 
\projectname organizes the workflow into four components: \textbf{Dataset Building}, \textbf{Dataset Attack}, \textbf{Detector Training}, and \textbf{Performance Evaluation}. 
It supports constructing custom benchmarks by generating MGT with configurable LLMs, applying 12 text attacks to test sets, training detectors via a unified interface, and reporting effectiveness, robustness, and efficiency. 
The platform provides both command-line and Web-based interfaces for user-friendly experimentation without code rewriting.
\end{abstract}

\begin{figure*}[!t]
  \centering
  \includegraphics[width=\textwidth]{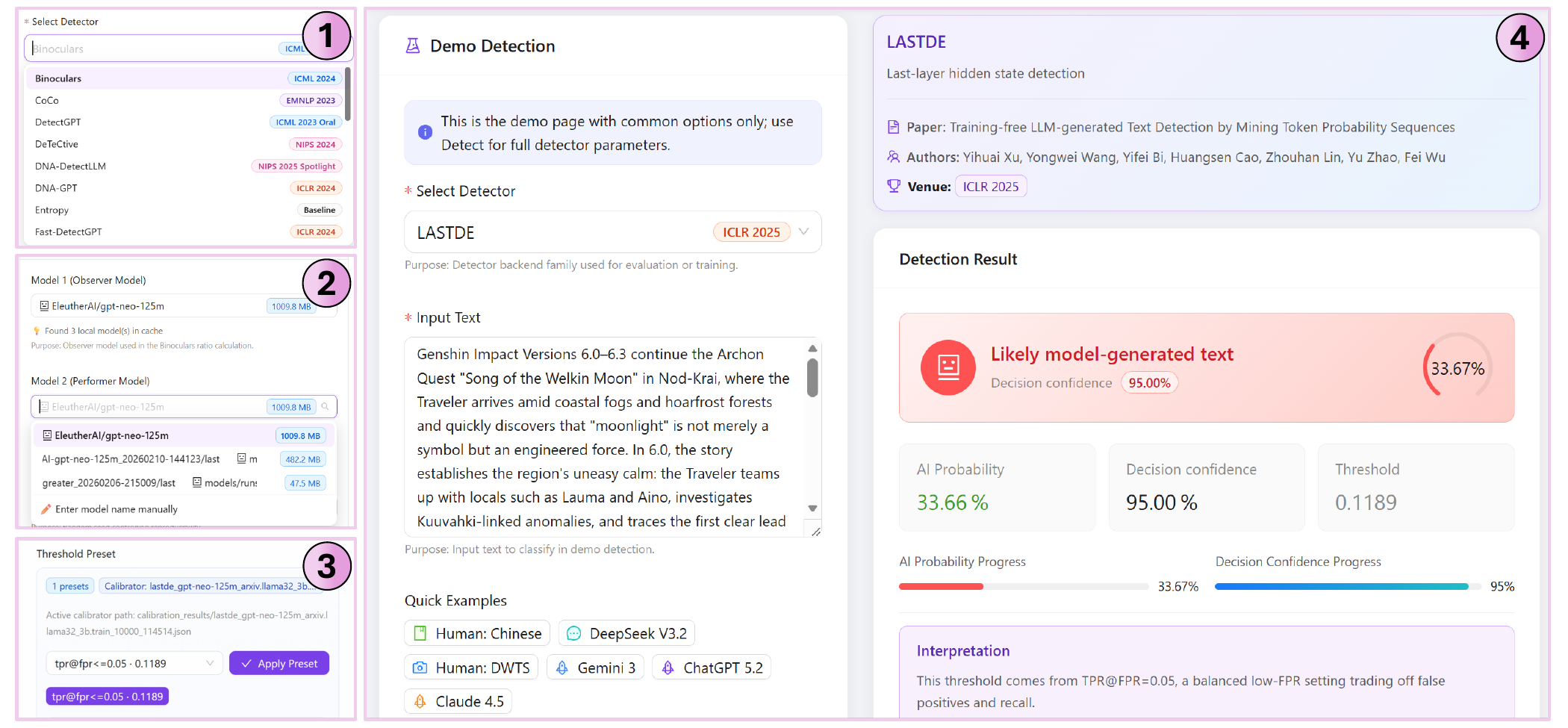}
  \caption{\textbf{Demo section of our \projectname.} Before detecting, users can select the detectors (subplot 1), and then select the models (subplot 2) and the parameters (subplot 3) used for detection. Then the users can input the text and run the detection (subplot 4), our detector will output the detection result (human-written or machine-generated) and the confidence of the result. Try our \projectname in \url{http://uncoverai.cn/}.}
  \label{fig:demo}
\end{figure*}
\input{sections/introduction}
\input{sections/related_work}

\input{sections/methodology}
\input{sections/experiment}
\input{sections/conclusion}


\bibliography{custom}

\clearpage

\appendix

\input{sections/appendixA}

\end{document}

%% file: sections/introduction.tex
\section{Introduction}
\label{sec:intro}

The rapid advancement of large language models (LLMs), such as ChatGPT~\citep{achiam2023gpt}, LLaMA~\citep{dubey2024llama}, and DeepSeek~\citep{liu2025deepseek}, has made it increasingly easy to generate fluent, human-like text at scale.
Although this capability enables a wide range of beneficial applications, it also lowers the barrier to producing misleading or malicious content, including fabricated news and sophisticated phishing messages.
To address these risks, researchers have developed various methods to detect machine-generated text (MGT)~\citep{su2023detectllm, hans2024spotting, bao2024fast, liu2024does, li2025iron}, aiming to distinguish LLM-generated content from human-written text and inform the users about the provenance of the textual contents.
However, the emerging MGT detectors lack a consistent platform for fair comparison under the same setting, leading to difficulty in evaluating the performance and robustness of the detector.


To address this gap, we develop \projectname, a unified platform for systematically evaluating the capabilities of existing MGT detectors with a user-friendly graphical interface.
\projectname not only supports document detection using selected detectors, similar to commercial products such as GPTZero\footnote{https://gptzero.me/}
 (see Figure~\ref{fig:demo} for demonstration), but also provides a fully customizable, full-cycle workflow for constructing and evaluating MGT detectors.
Specifically, it consists of four components: \textbf{Dataset Building}, \textbf{Dataset Attack}, \textbf{Detector Training}, and \textbf{Performance Evaluation}. 
The users are allowed to customize their dataset for training and testing by uploading human-written text corpora and configuring the generator for generating corresponding MGT. 
To support robustness evaluation, \projectname includes a \textbf{Dataset Attack} module which contains 12 common attack methods such as humanization~\cite{wang2024stumbling} and deletion~\cite{kukich1992techniques}, for perturbing the test set and providing a comprehensive testbed for assessing detector performance.
\textbf{Detector Training} module provides a configurable training pipeline for training the detectors.
To simplify the training process, we train 26 existing detectors using the standardized datasets from the Dataset Building module and release as off-the-shelf models.
In the \textbf{Performance Evaluation} component, we automatic the evaluation process and report the common metrics such as Accuracy, F1 score, AUROC, and TPR at FPR $=0.01$ and $0.001$.
Users can customize detector evaluation through either the command-line interface or a web-based interface without rewriting any code.


\projectname adopts a modular design, in which all detector and attack modules are managed through a registry mechanism. This design greatly improves extensibility and maintainability, enabling users to seamlessly add new methods, reuse existing components, and configure evaluation pipelines flexibly without modifying the core framework.
Our contributions are summarized as follows:
\begin{itemize}
    \item \textbf{Unified Evaluation Pipeline.} We present \projectname, a unified framework for MGT detector evaluation that standardizes the full workflow, including dataset construction, dataset attack, detector training, and performance evaluation. By unifying the data preparation, training, and evaluation process, \projectname provides a reproducible, configurable, and fair platform for comparison across detectors.
    

    \item \textbf{Comprehensive Benchmarking Suite.} \projectname supports systematic evaluation of 26 existing MGT detectors on both clean and attacked datasets, with comprehensive metrics such as Accuracy, F1, AUROC, and TPR at low FPR operating points. This design facilitates multi-dimensional analysis of detector performance, including effectiveness and robustness under text perturbations.

    \item \textbf{User-Friendly Extensible Framework.} \projectname provides both command-line and web-based interfaces, allowing users to customize evaluation pipelines without rewriting code. In addition, its modular, registry-based design for detector and attack modules improves extensibility and maintainability, making it easy to integrate new methods and support future research.
\end{itemize}

%% file: sections/related_work.tex
\section{Related Work}

\textbf{Machine-Generated Text (MGT) Detectors.} As LLMs continue to advance, a growing body of work has focused on detecting machine-generated text (MGT). Existing methods can be broadly categorized into \emph{metric-based} and \emph{model-based} detectors. Metric-based detectors~\cite{su2023detectllm,bao2024fast,hans2024spotting,xu2024training,zhu2025dna} require little or no supervised training and identify MGT using statistical cues, but they are often sensitive to adversarial perturbations and may yield poorly calibrated scores, making threshold selection difficult in real-world deployment. By contrast, model-based detectors~\cite{liu2022coco,hu2023radar,liu2024does,koike2024outfox,li2025iron} fine-tune text classifiers (e.g., BERT/RoBERTa) and generally achieve stronger detection performance, but they typically depend on substantial labeled data to reach high accuracy. \projectname integrates a wide range of advanced metric-based and model-based detectors under a unified framework for standardized evaluation and comparison.

\begin{figure*}[t!]
	\centering
	\resizebox{\textwidth}{!}{%
		\includegraphics{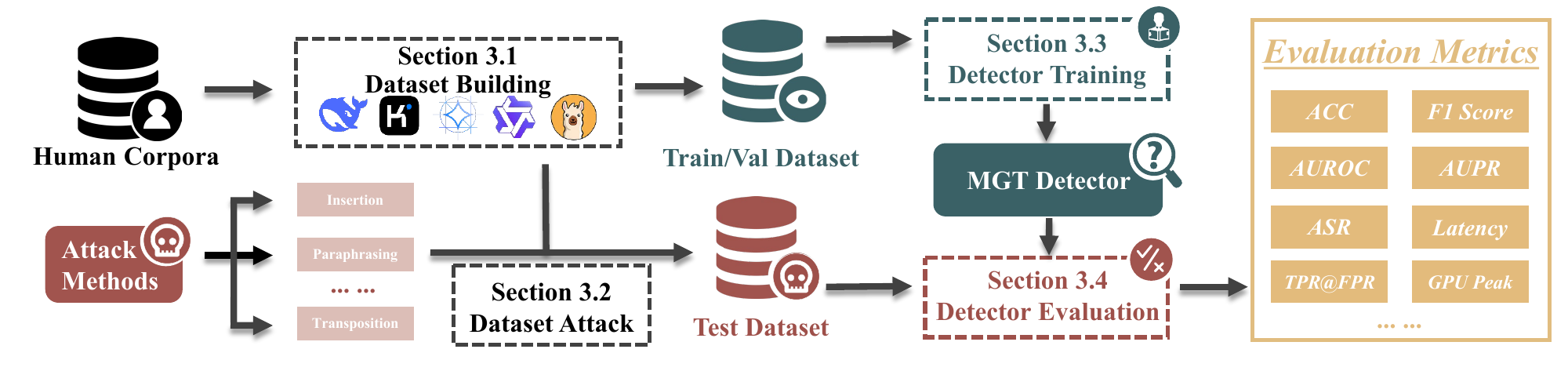} 
	}
	\caption{\textbf{Pipeline of our \projectname.} Users can use human corpora and configurable LLMs to build dataset (Section~\ref{sec:build}), and choose different attacks to generate attacked test dataset (Section~\ref{sec:attack}). Then users can use train/val dataset to train a detector (Section~\ref{sec:training}), test the detector in the test dataset, and obtain output metrics (Section~\ref{sec:evaluation}).}
	\label{fig:interface} 
\end{figure*}

\noindent\textbf{Machine-Generated Text (MGT) Benchmarks.}
With the rapid development of MGT detection, there have recently been several efforts toward building systematic evaluation frameworks for MGT detectors. Representative examples include \textbf{MGTBench}~\citep{he2024mgtbench}, \textbf{MGTBench 2.0}~\citep{liu2025generalization}, and \textbf{Stumbling Blocks}~\citep{wang2024stumbling}, which respectively advance benchmark standardization, academic-domain evaluation, and robustness stress testing under realistic attacks. These works provide valuable resources and evaluation protocols for analyzing detector performance from different perspectives. However, as shown in Table~\ref{tab:mgt_benchmark_compare_vertical}, existing frameworks still have limited support for unified end-to-end evaluation functionalities, such as integrated dataset construction, detector training, user-friendly front-end interaction, and comprehensive multi-metric reporting within a single system. In contrast, our \projectname is designed as an integrated and extensible framework that unifies dataset construction, attack generation, detector training, and detector evaluation for systematic MGT detector benchmarking.

\begin{table}[t]
\centering
\setlength{\tabcolsep}{4pt}
\renewcommand{\arraystretch}{1.15}
\resizebox{0.45\textwidth}{!}{%
\begin{tabular}{lcccc}
\toprule
\textbf{Feature} & \textbf{MGTBench} & \textbf{MGTBench 2.0} & \textbf{Stumbling Blocks} & \textbf{\projectname} \\
\midrule
\# Detectors          & 13 & 12 & 8  & \textbf{26} \\
\# Attacks            & 3  & 0  & 12 & \textbf{12} \\
\# Built-in Metrics   & -- & -- & 3 & \textbf{$\geq$5} \\
\midrule
Custom Dataset  & \color{red}\XSolidBrush & \color{red}\XSolidBrush & \color{red}\XSolidBrush & \color{red}\XSolidBrush \\
Custom Attacks & \color{red}\XSolidBrush & \color{red}\XSolidBrush & \color{red}\XSolidBrush & \color{red}\XSolidBrush \\
\midrule
Front-end UI          & \color{red}\XSolidBrush & \color{red}\XSolidBrush & \color{red}\XSolidBrush &\color{green}\Checkmark \\
CLI Support           & \color{green}\Checkmark & \color{green}\Checkmark & \color{green}\Checkmark & \color{green}\Checkmark \\
\bottomrule
\end{tabular}%
}
\caption{Comparison of features in MGTEval with existing MGT detectors evaluation framework.}
\label{tab:mgt_benchmark_compare_vertical}
\end{table}

%% file: sections/methodology.tex
\section{\projectname Framework}

In this section, we present the \projectname framework. \projectname consists of four components: \textbf{Build Dataset}, \textbf{Attack Dataset}, \textbf{Train Detector}, and \textbf{Evaluate Detector}. 
Our pipeline can be found in Figure~\ref{fig:interface}.
To make the pipeline extensible and reproducible, \projectname adopts a registry-based, configuration-driven design: all modules exchange a canonical record schema and save run manifests (configuration, random seed, and output artifacts), which reduces integration cost across new datasets and detectors.





\subsection{Dataset Building}
\label{sec:build}
The \textbf{Dataset Building} module standardizes heterogeneous corpora into a unified binary detection format and provides a reproducible pipeline for dataset loading and custom dataset construction.

\paragraph{Dataset Loading.}
\projectname supports loading datasets from local files (\texttt{.jsonl}, \texttt{.json}, \texttt{.csv}), directory-level recursive discovery, and iterable in-memory objects. In addition to flat binary records, it natively handles common MGT benchmark schemas such as HC3-style QA records and paired generation records, enabling users to seamlessly ingest both benchmark corpora and custom collections.
Supported dataset structures are summarized in Table~\ref{tab:mgteval_data_formats}, and \projectname covers widely used benchmarks including HC3~\cite{guo2023close}, SemEval 2024~\cite{semeval2024task8}, and M4~\cite{wang-etal-2024-m4}.

\paragraph{Custom Dataset Construction.}
Beyond directly loading existing datasets, \projectname enables \emph{user-defined} dataset construction by pairing human-labeled corpora with machine-generated counterparts produced by configurable LLM backends. Given a corpus consisting of human-written texts, users can specify one or multiple LLM generators and fully configure generation parameters, including prompt templates, decoding temperature, top-$k$/top-$p$ sampling, maximum length, and random seeds, to synthesize machine-written samples. The resulting dataset is automatically labeled (\texttt{label}=0 for human, \texttt{label}=1 for machine) and annotated with generation metadata (e.g., \texttt{model}, prompt ID, decoding settings), facilitating controlled benchmarking across generators and decoding regimes.

\begin{table}[t]
\centering
\small
\setlength{\tabcolsep}{4pt}
\renewcommand{\arraystretch}{1.12}
\resizebox{\columnwidth}{!}{%
\begin{tabular}{@{}p{3.1cm}p{9.4cm}@{}}
\toprule
\textbf{Format Type} & \textbf{Example Structure} \\
\midrule
Flat binary data &
[\{"text": "...", "label": 0/1, "id": "...", "source": "...", "lang": "...", "model": "..."\}] \\

HC3-style QA data &
[\{"human\_answers": ["..."], "chatgpt\_answers": ["..."], "source": "..."\}] \\

Paired generation data &
\{"original": [\{"text": "..."\}], "sample": [\{"text": "..."\}], "sampled": [\{"text": "..."\}], "rewritten": [\{"text": "..."\}]\} \\

Attack-aligned paired data &
\{"sample": [\{"text": "...", "attack": "..."\}], "meta": \{"base\_id": "...", "active\_attack": "..."\}\} \\

Standardized data &
\{"id": "...", "text": "...", "label": 0/1, "source": "...", "lang": "...", "model": "...", "attack": "..."\} \\
\bottomrule
\end{tabular}%
}
\caption{Supported dataset structures in \projectname.}
\label{tab:mgteval_data_formats}
\end{table}

\subsection{Dataset Attack}
\label{sec:attack}
After dataset construction, \textbf{Dataset Attack} is used to build adversarial evaluation sets for robustness analysis. The key objective is to stress-test detectors under realistic evasion perturbations while preserving the human reference distribution.

Given either (i) datasets generated by \textbf{Dataset Building} or (ii) any external dataset supported by the unified loader, \projectname first aligns all records to the standard binary format. As is common practice in prior work~\cite{wang2024stumbling}, it then attacks \emph{only} machine-generated samples (\texttt{label}=1), while keeping human samples unchanged. 
This setting avoids unnecessary human-text distortion and ensures a clean robustness comparison between clean and attacked machine outputs.

For each eligible machine sample, the module generates one or multiple attacked variants and stores them with provenance metadata (e.g., sample identity and attack type), so that paired and traceable robustness evaluation can be performed downstream.


\subsection{Detector Training}
\label{sec:training}
Using the aligned training split from \textbf{Build Dataset}, \projectname supports both metric-based and model-based detector training under a unified interface.

\paragraph{Metric-based Detectors.}
For metric-based detectors, the model output is a scalar score rather than a calibrated probability. Therefore, \projectname treats calibration as the training step for this detector family. Given a training set \(\{(x_i, y_i)\}_{i=1}^{N}\), where \(y_i \in \{0,1\}\), we first compute detector scores \(s_i=g(x_i)\). We then fit a one-dimensional logistic regression to map scores into probabilities:
\begin{equation}
p_i=\sigma(\alpha s_i+\beta)=\frac{1}{1+\exp\!\left(-(\alpha s_i+\beta)\right)},
\label{eq:metric_prob_map}
\end{equation}
where \(\alpha,\beta\) are learned by minimizing binary cross-entropy:
\begin{equation}
\small
\begin{aligned}
(\alpha^\ast,\beta^\ast)
&=
\arg\min_{\alpha,\beta}
\Biggl[
-\frac{1}{N}\sum_{i=1}^{N}
\Bigl(
y_i\log p_i \\
&\qquad\qquad\qquad\quad + (1-y_i)\log(1-p_i)
\Bigr)
\Biggr].
\end{aligned}
\label{eq:metric_calib_obj}
\end{equation}
This mapping enables unified thresholding and fair comparison across heterogeneous metric detectors. \projectname also exposes detector-specific controls, such as scoring model choice, perturbation settings (e.g., perturbation number/strength), calibration dataset path, threshold policy, and training sample size (e.g., \texttt{sample\_k}, random seed).

\paragraph{Model-based Detectors.}
For model-based detectors, \projectname performs end-to-end supervised optimization on the unified binary dataset. These detectors directly output decision-ready logits/probabilities, so no separate score-to-probability fitting is required. The framework supports configurable optimization and data-scale settings, including backbone/model selection, learning rate, batch size, number of epochs, maximum sequence length, weight decay, warmup ratio, gradient accumulation, precision/device options, train/validation/test split ratios, and explicit subsampling for controlled data-budget experiments.  
Detectors that are inference-only (training-free) are skipped in this stage and evaluated directly in the next stage.

\subsection{Performance Evaluation}
\label{sec:evaluation}
After training (or direct loading for training-free detectors), \projectname evaluates detectors on held-out test sets, and optionally on attacked test sets generated by \textbf{Dataset Attack}. We organize evaluation into three aspects: \emph{effectiveness}, \emph{robustness}, and \emph{efficiency}.

For protocol consistency, threshold-dependent metrics use a fixed threshold determined during the training/calibration stage, and the same threshold is reused for both clean and attacked evaluation of the same detector. This design ensures fair comparison across settings and avoids threshold re-tuning on attacked data.

\paragraph{Effectiveness Metrics}
To measure detection quality on clean test sets (and optionally attacked test sets), \projectname reports a set of standard classification and ranking metrics.
Specifically, we report \textbf{Accuracy (ACC)} for overall correctness, \textbf{Precision}, \textbf{Recall}, and \textbf{F1} for positive-class detection quality, and threshold-free ranking metrics including \textbf{AUROC} and \textbf{AUPR}.
To characterize performance under strict deployment constraints, we also report low-false-positive operating-point metrics such as \textbf{TPR@FPR=\(\alpha\)} (e.g., \(\alpha=0.01\) and \(0.001\)), which are particularly important in real-world moderation and integrity scenarios.

\paragraph{Robustness Metrics}
To evaluate robustness against evasive modifications, \projectname supports attacked test sets generated by the \textbf{Attack Dataset} module and reports robustness-oriented metrics on these sets.
In particular, we use \textbf{Attack Success Rate (ASR)} to quantify how often attacks can flip detector decisions on samples that were originally classified correctly in the clean setting.
This paired evaluation protocol enables traceable robustness analysis across attack types and detector families.

\begin{table*}[t]
\centering
\small
\setlength{\tabcolsep}{4pt}
\renewcommand{\arraystretch}{1.08}
\resizebox{\textwidth}{!}{%
\begin{tabular}{ccccccccc}
\toprule
\textbf{Detector} & \textbf{Accuracy (\%, $\uparrow$)} & \textbf{F1 (\%, $\uparrow$)} & \textbf{AUROC (\%, $\uparrow$)} & \textbf{TPR@FPR=0.01 (\%, $\uparrow$)} & \textbf{TPR@FPR=0.001 (\%, $\uparrow$)} & \textbf{Avg Latency (ms/sample) ($\downarrow$)} & \textbf{GPU Peak (GiB) ($\downarrow$)} \\
\midrule
Binoculars~\cite{hans2024spotting}         & 92.00 & 91.75 & 95.84 & 83.00 & 81.00 & 21.85 & 3.64 \\
CoCo~\cite{liu2022coco}               & 97.50 & 97.56 & \underline{99.95} & \underline{99.00} & \underline{98.00} & 73.13 & 0.63 \\
DetectGPT~\cite{mitchell2023detectgpt}          & 50.00 & 66.44 & 62.99 & 1.00 & 1.00 & 585.28 & 2.92 \\
DeTeCtive~\cite{guo2024detective}          & 70.50 & 67.76 & 80.69 & 23.00 & 23.00 & 42.67 & 1.82 \\
DNA-DetectLLM~\cite{zhu2025dna}      & 89.50 & 89.01 & 95.42 & 80.00 & 80.00 & 31.03 & 3.26 \\
DNA-GPT~\cite{yang2023dna}            & 79.00 & 80.00 & 88.83 & 23.00 & 19.00 & 295.89 & 5.59 \\
Entropy~\cite{gehrmann2019gltr}            & 72.50 & 69.61 & 18.02 & 0.00 & 0.00 & 16.80 & 0.50 \\
Fast-DetectGPT~\cite{bao2024fast}     & 90.50 & 90.26 & 95.61 & 85.00 & 80.00 & 13.64 & 1.04 \\
GLTR~\cite{gehrmann2019gltr}                & 90.00 & 90.20 & 94.01 & 20.00 & 1.00 & 26.69 & 1.01 \\
GREATER~\cite{li2025iron}             & 97.50 & 97.46 & 99.92 & 96.00 & 96.00 & 7.01 & \textbf{0.14} \\
RoBERTa-Base~\cite{liu2019roberta}        & 93.50 & 93.90 & 99.72 & 87.00 & 87.00 & 6.91 & \underline{0.32} \\
LASTDE~\cite{xu2024training}              & 88.50 & 88.21 & 94.94 & 32.00 & 3.00 & 14.99 & 0.41 \\
LASTDE++~\cite{xu2024training}                 & 91.50 & 91.10 & 96.24 & 84.00 & 84.00 & 63.10 & 1.19 \\
Likelihood~\cite{gehrmann2019gltr}               & 87.00 & 86.87 & 94.06 & 30.00 & 0.00 & 14.59 & 0.41 \\
LogRank~\cite{gehrmann2019gltr}            & 88.00 & 87.88 & 94.73 & 36.00 & 0.00 & 18.03 & 1.03 \\
Longformer~\cite{beltagy2020longformer}         & \textbf{99.50} & \textbf{99.50} & \textbf{99.99} & \textbf{100.00} & \textbf{99.00} & 20.29 & 0.83 \\
LRR~\cite{su2023detectllm}                & 85.50 & 84.82 & 95.13 & 47.00 & 43.00 & 18.86 & 1.01 \\
MPU~\cite{tian2023multiscale}                & 90.50 & 91.32 & 99.88 & 98.00 & \underline{98.00} & \underline{6.87} & \underline{0.32} \\
NPR~\cite{su2023detectllm}                 & 50.00 & 66.67 & 80.04 & 3.00 & 0.00 & 1132.42 & 0.96 \\
OpenAI Detector    & 73.50 & 65.81 & 81.25 & 41.00 & 39.00 & 959.82 & \underline{0.32} \\
PECOLA~\cite{liu2024does}             & 94.50 & 94.74 & 99.69 & \underline{99.00} & 60.00 & \textbf{6.86} & \underline{0.32} \\
RADAR~\cite{hu2023radar}              & 26.50 & 29.67 & 15.65 & 0.00 & 0.00 & 2067.45 & 0.76 \\
RAiDAr~\cite{mao2024raidar}             & \underline{98.00} & \underline{97.98} & 99.63 & 97.00 & 86.00 & 427.40 & 14.99 \\
Rank~\cite{gehrmann2019gltr}                & 77.50 & 78.47 & 85.69 & 33.00 & 5.00 & 14.87 & 1.03 \\
SimpleAI Detector~\cite{guo2023close} & 56.00 & 27.87 & 60.99 & 12.00 & 9.00 & 611.88 & \underline{0.32} \\
TOCSIN~\cite{ma2024zero}             & 59.00 & 70.50 & 95.03 & 79.00 & 0.00 & 110.73 & 3.73 \\
\bottomrule
\end{tabular}%
}
\caption{\textbf{Detector performance summary.} Avg Latency denotes the average detection time per sample during evaluation, measured in milliseconds (ms/sample), and GPU Peak denotes the peak GPU memory consumption during evaluation (GiB). Best values are in bold and second-best values are underlined.}
\label{tab:detector_summary}
\end{table*}

\paragraph{Efficiency Metrics}
To measure practical usability, \projectname reports runtime-oriented metrics during detector evaluation.
These include \textbf{evaluation time} (end-to-end runtime for processing a test set), \textbf{throughput} (samples processed per second), and \textbf{latency per sample} (average inference cost per sample).
Together, these metrics help users compare not only detection quality but also deployment efficiency under different detector architectures and configurations.

\paragraph{Output and Reporting}
Beyond scalar summary metrics, \projectname can export structured evaluation results for downstream analysis, including detector-level summaries and (when available) sample-level predictions with metadata such as source, language, generator, and attack type.
This supports grouped/sliced evaluation and facilitates reproducible comparison across datasets, attacks, and detectors.

\section{Interface Design}
\label{sec:interface}
The \projectname WebUI is designed as a task-oriented control panel that mirrors the end-to-end workflow of the framework while minimizing configuration friction for non-expert users, an overview can be found in Figure~\ref{fig:interface}. The interface organizes core operations into four pages consistent with the backend pipeline: \textit{Build Dataset} (Figure~\ref{fig:interface} Part 1), \textit{Attack Dataset} (Figure~\ref{fig:interface} Part 2), \textit{Train Detector} (Figure~\ref{fig:interface} Part 3), and \textit{Evaluate Detector} (Figure~\ref{fig:interface} Part 4). Each page provides structured parameter forms with default presets, inline help, and validation to prevent invalid runs. To support reproducibility, the UI exposes seed, data path, and output path controls, and synchronizes with configuration files used by the CLI. During execution, users can monitor real-time logs, job status, and resource-related metadata, and directly access exported artifacts such as JSON/CSV reports and plots. In addition to the core pipeline pages, the WebUI includes an extra \textit{Demo} section (Figure~\ref{fig:demo}) for interactive single- or small-batch testing, where users can quickly inspect detector outputs and compare clean versus attacked text behavior before launching full-scale experiments.
More screenshots of interface can be found in Appendix~\ref{sec:screenshots}.

%% file: sections/experiment.tex
\section{Experiments and Results}
\label{sec:exp}

\noindent\textbf{Experiment Setting.}
We evaluate \projectname on a clean held-out test set constructed by the \textbf{Build Dataset} module.
Specifically, human-written texts are sampled from the SemEval 2024 dataset~\citep{semeval2024task8}, while machine-generated texts are produced by \texttt{Qwen3-4B-2507-Instruct}~\cite{yang2025qwen3}.
The final dataset contains 2{,}000 samples in total, with a balanced human-to-machine ratio of 1:1.
We split the dataset into training/validation/test sets with a ratio of 8:1:1 and report results on the test split.
For detector implementation consistency in \projectname, model-based detectors use \texttt{RoBERTa-Base}~\cite{liu2019roberta} as the backbone, while metric-based detectors use \texttt{GPT-Neo-125M}~\cite{gao2020pile} as the scoring model and \texttt{t5-small}~\cite{2020t5} as the perturbation model (for methods requiring perturbation-based scoring).
All detectors are evaluated under the same pipeline in \projectname, and we report effectiveness, robustness, and efficiency metrics in Table~\ref{tab:detector_summary}.

\noindent\textbf{Experiment Results.}
We report the performance of 26 detectors in Table~\ref{tab:detector_summary} and draw the following observations.
\textbf{(i) Strong overall performance under a unified protocol.}
\texttt{Longformer} achieves the best results on all major effectiveness and low-FPR robustness metrics, reaching 99.50\% Accuracy, 99.50\% F1, 99.99\% AUROC, 100.00\% TPR@FPR=0.01, and 99.00\% TPR@FPR=0.001.
Compared with the second-best detector in Accuracy/F1 (\texttt{RAiDAr}), it improves by +1.50 pp Accuracy and +1.52 pp F1, indicating a clear margin under the same data and evaluation settings.
\textbf{(ii) Clear efficiency--effectiveness trade-offs across detector families.}
While \texttt{Longformer} provides the strongest overall detection quality, it is not the most efficient method.
The fastest detector is \texttt{PECoLA} (6.86 ms/sample), followed closely by \texttt{MPU} (6.87 ms/sample), \texttt{RoBERTa-Base} (6.91 ms/sample), and \texttt{GREATER} (7.01 ms/sample).
Among these efficient detectors, \texttt{GREATER} is particularly attractive because it combines near-SOTA performance (97.50\% Accuracy, 99.92\% AUROC, 96.00\%/96.00\% TPR at FPR=0.01/0.001) with the lowest GPU peak memory usage (0.14 GiB), substantially lower than the second-best memory tier (0.32 GiB).
In contrast, some detectors (e.g., \texttt{RAiDAr}) achieve strong accuracy but incur much higher computational cost (427.40 ms/sample and 14.99 GiB GPU peak), highlighting the importance of reporting efficiency metrics together with detection quality.
\textbf{(iii) Multi-metric evaluation is necessary to reveal calibration and threshold sensitivity.}
Several detectors exhibit large discrepancies between threshold-free and threshold-dependent metrics.
For example, \texttt{TOCSIN} obtains a strong AUROC (95.03\%) but much lower Accuracy (59.00\%) and a TPR@FPR=0.001 of 0.00\%, suggesting sensitivity to thresholding and operating-point selection.
Similarly, \texttt{DetectGPT} and \texttt{NPR} are substantially slower (585.28 and 1132.42 ms/sample, respectively) while delivering limited Accuracy gains (both 50.00\% Accuracy in this setting).
We also observe severe degradation for some methods (e.g., \texttt{RADAR} and \texttt{Entropy}), which further demonstrates the value of \projectname as a unified benchmark for systematically comparing detector effectiveness, robustness, and efficiency under identical protocols.

%% file: sections/conclusion.tex
\section{Conclusion}
We introduced \projectname, a unified framework for systematic evaluation of machine-generated text detectors. Unlike prior benchmark efforts that focus on specific datasets, tasks, or robustness settings, \projectname provides an end-to-end evaluation pipeline that unifies dataset construction, attack generation, detector training/calibration, and detector evaluation under a consistent protocol. The framework integrates 26 detectors and 12 attacks, supports comprehensive metrics covering effectiveness, robustness, and efficiency, and offers both CLI and WebUI interfaces to improve usability and reproducibility. 
\projectname is designed as a modular, registry-based infrastructure to facilitate future extensions. In future work, we plan to expand support for multilingual and cross-domain benchmarks, richer attack compositions, attribution-oriented tasks, and broader detector families (including newer LLM- and API-based detectors), further advancing standardized and reproducible MGT detection evaluation.

\section*{Limitation}
Despite the practical utility and comprehensive design of \projectname for systematic MGT detector evaluation, it still exhibits several limitations:  
i) Although \projectname unifies dataset construction, attack generation, detector training, and evaluation under a single framework, running the full pipeline (especially for multiple detectors and attacks) can be computationally expensive and time-consuming, which may limit usability in resource-constrained environments.  
ii) While \projectname integrates a broad set of detectors and attacks through a registry-based modular design, some detector-specific implementation details or original hyperparameter settings may not be fully preserved under a unified protocol, which can lead to discrepancies from results reported in the original papers.  
iii) The current version of \projectname mainly focuses on binary MGT detection (human vs.\ machine) and a fixed set of text-level attacks. As a result, it does not yet fully cover more complex settings, such as fine-grained source attribution, multilingual large-scale benchmarking, or adaptive attack--defense co-evaluation, which may constrain its general applicability in broader real-world scenarios.

\section*{Acknowledgment}
This work is supported by National Natural Science Foundation of China (62272371, 62103323) and Fundamental Research Funds for the Central Universities under grant (xzy012024144, xzy012025043). The author Xiaoming Liu gratefully acknowledges the support of K. C. Wong Education Foundation.

\section*{Ethics Statement}
\label{sec:ethics}

This work presents \projectname, a benchmarking and evaluation framework for machine-generated text (MGT) detectors. Our goal is to improve the rigor, reproducibility, and transparency of detector evaluation, which may support safer deployment of MGT detection systems in applications such as content moderation, platform integrity, and academic integrity analysis.

\paragraph{Potential benefits.}
By standardizing dataset construction, attack generation, training/calibration, and evaluation, \projectname reduces protocol inconsistency and helps the community identify detector strengths and failure modes more reliably. We believe this can improve scientific comparability and encourage more robust detector development.

\paragraph{Dual-use risks.}
The framework includes attack-generation functionality (\textbf{Attack Dataset}) for robustness evaluation. Such functionality could potentially be misused to study detector weaknesses for evasion purposes. We include attacks for defensive benchmarking and stress testing, not to facilitate harmful misuse. In practice, we recommend responsible release and use of attack modules, including clear documentation, controlled benchmarking settings, and avoiding claims that attacks guarantee bypass in real systems.

\paragraph{Risk of false positives and deployment harm.}
MGT detectors can make incorrect predictions, including false positives on human-written text and false negatives on machine-generated text. These errors may cause harm if detector outputs are used as sole evidence in high-stakes decisions (e.g., academic penalties, moderation sanctions, or fraud accusations). \projectname is an evaluation framework, not a decision policy. We recommend that detector outputs be used as one signal among others, with human oversight in sensitive applications.

\paragraph{Data and privacy considerations.}
\projectname is designed to operate on user-provided or publicly available corpora. Users remain responsible for ensuring that the datasets they load, generate, or export comply with applicable licenses, privacy requirements, and institutional policies. When evaluating on sensitive data, users should adopt appropriate anonymization and access-control practices.

\paragraph{Bias and generalization.}
Detector performance may vary across languages, domains, and writing styles, and benchmarking results may reflect biases in the chosen datasets and generators. We encourage users to evaluate detectors on diverse datasets and report subgroup performance whenever possible.

Overall, we view \projectname as infrastructure for more transparent and reproducible MGT detector evaluation, and we encourage its use in ways that prioritize safety, fairness, and responsible deployment.


%% file: sections/appendixA.tex
\clearpage

\section{More Screenshots of \projectname}
\label{sec:screenshots}
\begin{figure*}[t]
	\centering
	\resizebox{1\textwidth}{!}{%
		\includegraphics{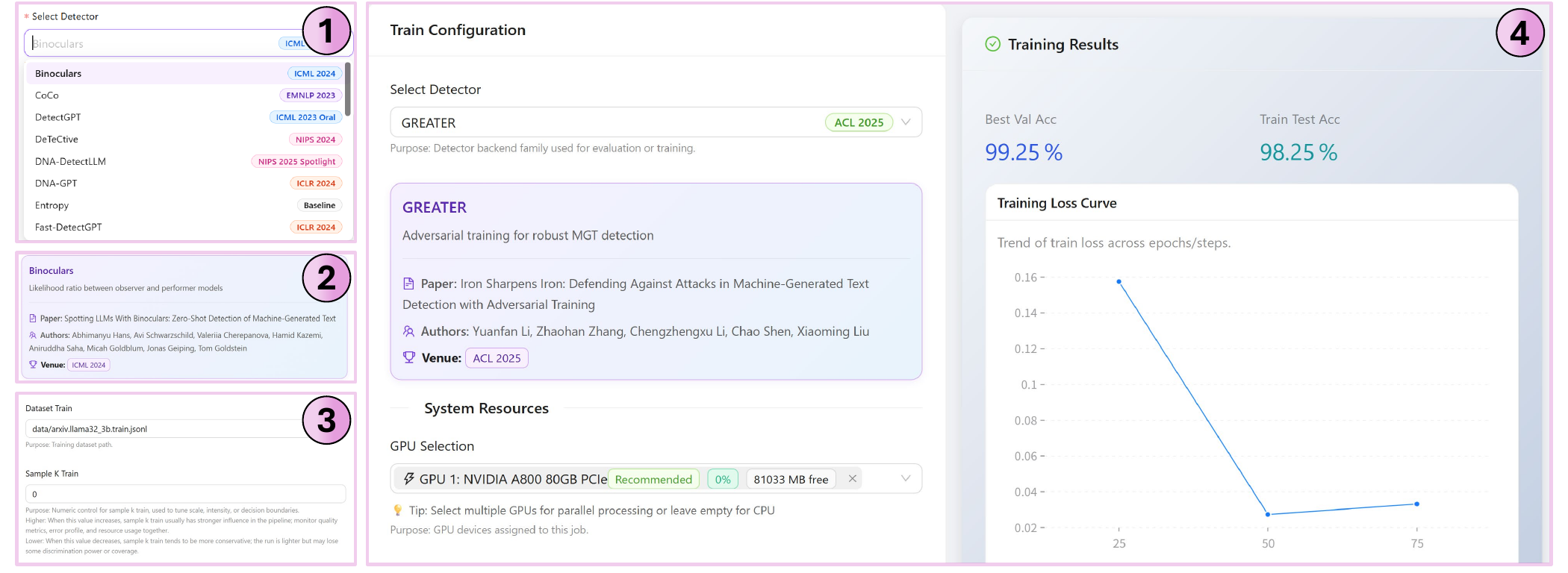} 
	}
	\caption{\textbf{The \emph{Detector Training} Page.}
    Users are allowed to select the detector to train from the available options (subplot 1) and access a concise summary of its metadata, including a high-level description, the corresponding paper, and its publication venue (subplot 2).
    The interface further allows users to configure training-related settings, such as the choice of training dataset and the number of samples to be used (subplot 3).
    After training, the system presents the training results, including evaluation accuracy on the validation set and, when available, on the test set, as well as the loss trajectory recorded throughout the training process (subplot 4).}
	\label{fig:train} 
\end{figure*}

\begin{figure*}[t]
	\centering
	\resizebox{1\textwidth}{!}{%
		\includegraphics{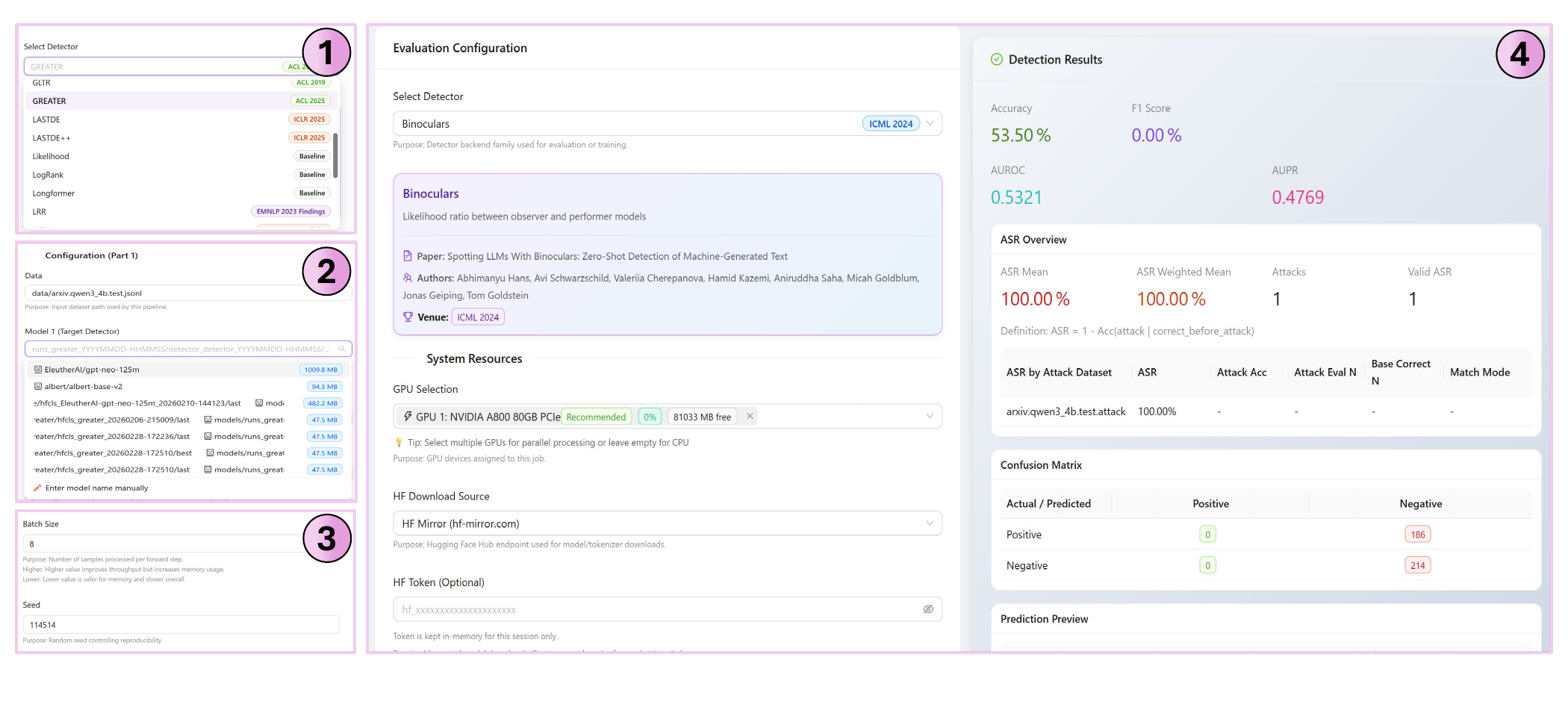} 
	}
	\caption{\textbf{The \emph{Performance Evaluation} Page.} 
    Users are allowed to select a detector to evaluate from the available options (subplot 1) and choose the existing evaluation dataset and checkpoint to be used (subplot 2).
    The interface also allows configuration of evaluation parameters, such as batch size and random seed (subplot 3). 
    Once the evaluation is completed, the system presents a comprehensive set of results, including Accuracy, F1 score, AUROC, AUPR, ASR, the confusion matrix, and additional metrics (subplot 4).
    }
	\label{fig:detect} 
\end{figure*}

Figure~\ref{fig:build}--Figure~\ref{fig:detect} provides additional screenshots of \projectname. Users can build custom datasets by uploading human-written texts and selecting an LLM to generate MGT (Figure~\ref{fig:build}), optionally applying attacks to create an attacked test set. In the training page (Figure~\ref{fig:train}), users can choose datasets and supported MGT detectors to train a model, while the evaluation page (Figure~\ref{fig:detect}) reports holistic metrics for assessing performance.


\vspace{0.2cm}
\begin{strip}
    \centering
    \includegraphics[width=\textwidth]{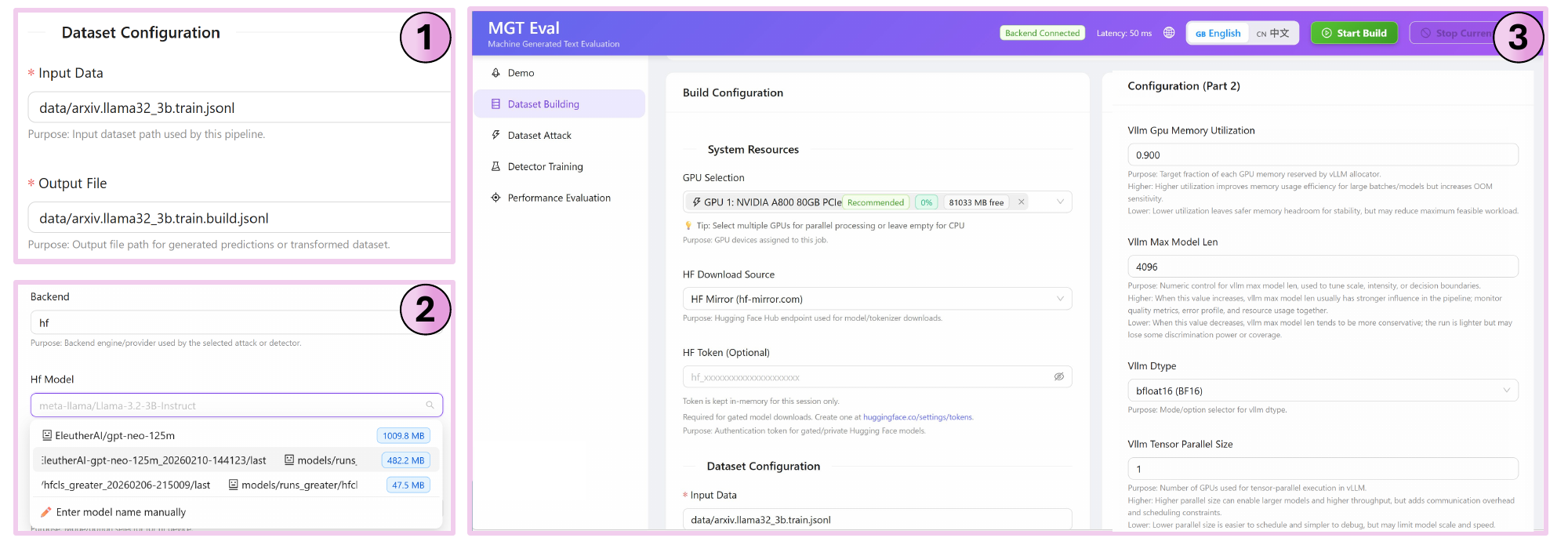}
    \captionof{figure}{\textbf{The \emph{Dataset Building} Page. }
    The users are allowed to specify the input path for human-written texts and the output directory where the constructed dataset will be saved (subplot 1). 
    Users can also select the LLM used to generate machine-generated texts (MGTs) that mimic the uploaded human samples (subplot 2).
    This page also provides additional configurable options, including the LLM temperature, maximum output tokens, and sampling hyperparameters such as Top-k and Top-p. (subplot 3) 
}
    \label{fig:build}
\end{strip}

\noindent
\begin{minipage}{\textwidth}
    \centering
    \includegraphics[width=\textwidth]{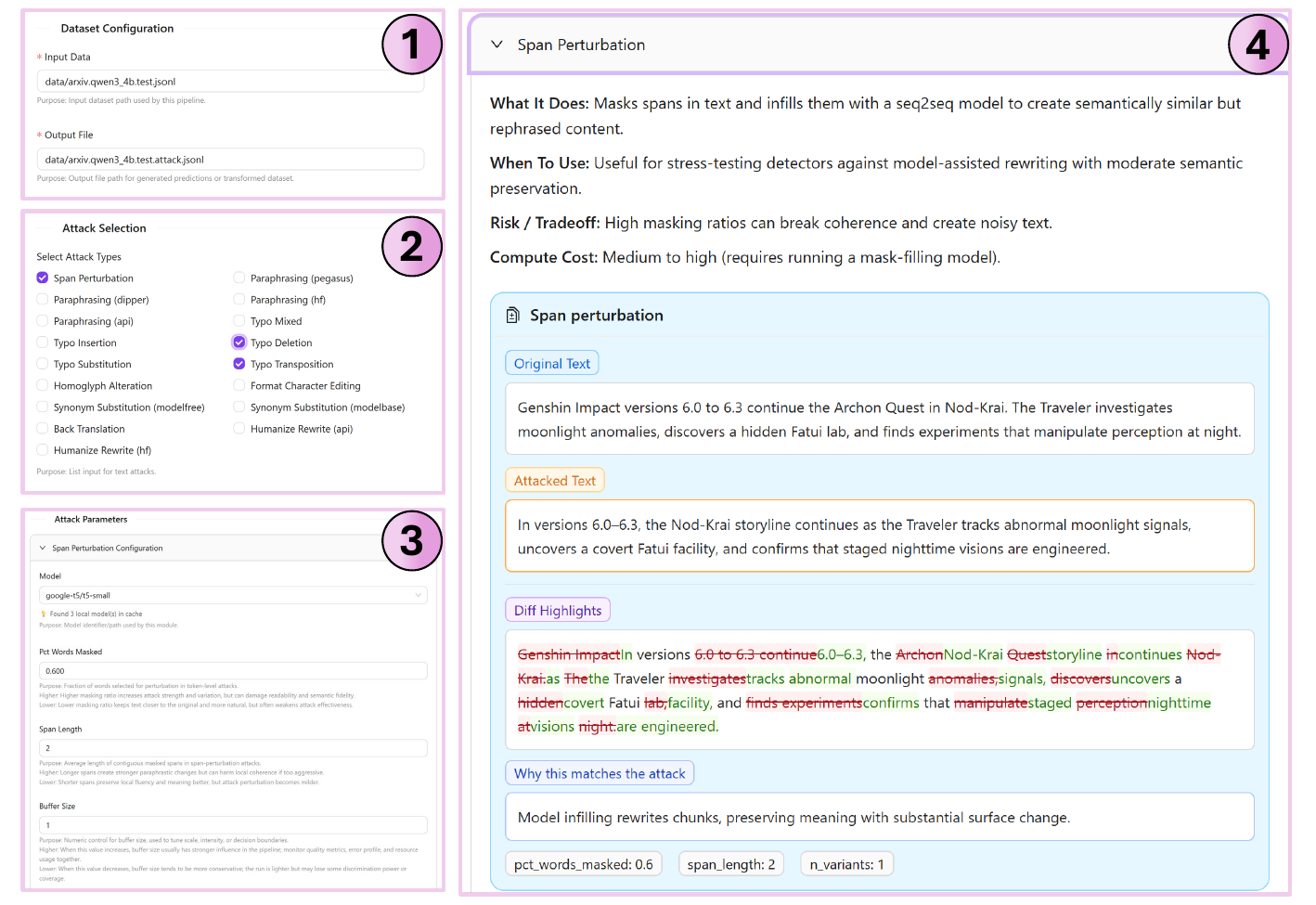}
    \captionof{figure}{\textbf{The \emph{Dataset Attack} Page.} Users are allowed to select the dataset to be attacked as the Input Data and specify the output directory for the attacked dataset (subplot 1). Users apply multiple attack methods to the dataset by selecting the supported attacks (subplot 2). For each selected attack, users adjust method-specific parameters, for example, controlling the proportion of tokens to be modified (subplot 3). To facilitate informed configuration, the page also presents descriptions and usage examples for all available attack methods (subplot 4).
}
    \label{fig:build}
\end{minipage}


